\documentstyle[prl,aps,epsf]{revtex}

\begin{document}
\def \beq {\begin{equation}}
\def \eeq {\end{equation}}
\def \bes {\begin{eqnarray}}
\def \ees {\end{eqnarray}}
\def \ni {\noindent}
\def \nn {\nonumber}
\def \z {\tilde{z}}
\def \kp {k_{\bot}}
\def \e {\varepsilon}
\title{Thermodynamical aspects of the Casimir force between
real metals at nonzero temperature}

\author{
V.~B.~Bezerra,
G.~L.~Klimchitskaya\footnote{On leave from
North-West Polytechnical University,
\\St.Petersburg, Russia.
E-mail: galina@fisica.ufpb.br}
and V.~M.~Mostepanenko\footnote{On leave from
Research and Innovation Enterprise \\ ``Modus'',
Moscow, Russia.
E-mail: mostep@fisica.ufpb.br}
}
\address
{Departamento de F\'{\i}sica, Universidade Federal da Para\'{\i}ba,
C.P.~5008, CEP 58059-970,
Jo\~{a}o Pessoa, Pb-Brazil
}
\maketitle

{\abstract{We investigate the thermodynamical aspects of
the Casimir effect in the case of plane parallel plates
made of real metals. The thermal corrections to the Casimir
force between real metals were recently computed by several
authors using different approaches based on the Lifshitz formula
with diverse results. Both the Drude and plasma models were
used to describe a real metal. We calculate the entropy density
of photons between metallic plates as a function of the surface
separation and temperature. Some of these approaches are
demonstrated to lead to negative values of entropy and to
nonzero entropy at zero temperature depending on the parameters
of the system. The conclusion is that these approaches are
in contradiction with the third law of thermodynamics and
must be rejected. It is shown that the plasma dielectric function
in combination with the unmodified Lifshitz formula is in perfect
agreement with the general principles of thermodynamics. As to the
Drude dielectric function, the modification of the zero-frequency
term of the Lifshitz formula is outlined that not to violate
the laws of thermodynamics.
}}


\section{Introduction}

Recent advances in experimental investigation of the Casimir effect
(see Refs.~\cite{1,2,3,4,5,6,7,8,8a,8b} and the review \cite{9}) have 
given impetus to extensive theoretical studies of different corrections
to the Casimir force. Casimir force arises between two closely
spaced neutral bodies due to the existence of zero-point electromagnetic
fluctuations. It is one of the rare macroscopic manifestations of
quantum phenomena. For this reason, it received widespread attention.
Moreover, currently the Casimir effect finds applications in fundamental
physics for constraining hypothetical forces predicted by different
extensions to the standard model \cite{10,11,12} and also in
nanotechnology \cite{7,8}.

Originally the Casimir force was computed between two infinitely large 
plane parallel plates made of ideal metal \cite{13}. Corrections to this
ideal result are caused by the geometrical factors (restricted area of the
plates and surface roughness), finite conductivity of the boundary metal
and nonzero temperature. Geometrical factors are detailly examined in the
literature (see, e.g., their study in Refs.~\cite{14,15,16,17,18} and also
in \cite{9}). Corrections caused by the finite conductivity of a metal
\cite{19,20,21,22,23,24} and by nonzero temperature \cite{20,25,26},
when considered separately, also received much attention and wholly
satisfactory results were obtained (see also \cite{9}).

For experimental purposes the combined effect of different corrections
to the Casimir force was found to be of large importance. The effect of
surface roughness, combined with any other corrections, can be effectively
computed by the method of geometrical averaging \cite{9,17}. So one comes
face to face with the problem how to find the combined action of finite
conductivity and nonzero temperature onto  the Casimir force. At first
glance it would seem that there is an easy way to solve this problem. Use
could be made of the famous Lifshitz formula \cite{19,27} for the Casimir
force at nonzero temperature acting between two dielectric semispaces
by the substitution of the dielectric permittivity function describing
real metals (on the basis of the plasma model, Drude model or optical
tabulated data for the complex refractive index). This was done recently 
by different authors \cite{28,29,30,31,32,33,34,35,36} and unexpectedly 
led to conflicting results.

In Refs.~\cite{30,31,33} the corrections to the Casimir force due to the 
combined effect of the finite conductivity and nonzero temperature were
calculated in the framework of the Lifshitz formula and of the free electron
plasma model. The obtained results are in agreement. Temperature
corrections are positive and smoothly transform to those for an ideal
metal in the limit of infinite plasma frequency.

In Refs.~\cite{28,29} the Drude dielectric function was substituted into
the Lifshitz formula. The temperature corrections to the Casimir force
were found to be negative at small space separations between the plates.
The asymptotic values of the Casimir force at high temperature (large
separations) were found to be two times smaller than for ideal metal
irrespective of how high the conductivity of real metal is. Thus, the
results of \cite{28,29} do not convert smoothly to the results, given
by the plasma model, when the relaxation goes to zero and also to the
results, found for an ideal metal, when the plasma frequency goes to
infinity. So extraordinary properties of the obtained results were
attempted to explain in \cite{28,29} by the principal role of nonzero
relaxation. Mathematically these properties are caused by the zero value 
of the reflection coefficient at zero frequency as given by the 
Lifshitz formula
for photons with perpendicular polarization in the framework of the Drude
model.

In Refs.~\cite{35,36} both the plasma and the Drude models were used
supplemented by the special prescription modifying the zero-frequency
term of the Lifshitz formula in the same way as 
was done in Ref.~\cite{20} for
the case of an ideal metal. As a result, large temperature 
corrections arise to the Casimir force at small separations that are linear
in temperature. At large separations (high temperatures), asymptotic
of the Casimir force in \cite{35,36} do not demonstrate any finite
conductivity correction starting from a separation of several micrometers.

As explained in Refs.~\cite{32,33}, actually the Drude model is outside
of the application range of the Lifshitz formula and, specifically,
the zero-frequency term of this formula must be modified in an appropriate
way in order to incorporate the dissipative media. This was demonstrated on 
the basis of a new derivation of the Lifshitz formula \cite{9} in the
framework of quantum field theory in Matsubara formulation. 
In Ref.~\cite{32} the new prescription for the zero-frequency term of the
Lifshitz formula was proposed which is not subject to the above-mentioned
disadvantages (see also \cite{34}).
Discussions regarding the correct description of the thermal Casimir
force between real metals are, however, being continued (see recent
Comments \cite{37,38} and Replies \cite{39,40} supporting just the 
opposite points
of view). The necessity to conclusively resolve a problem is apparent
when it is considered that the experiment is already nearing the registration
of the thermal corrections to the Casimir force.

In the present paper we analyse the thermal Casimir force between real metals
on the basis of the fundamental principles of thermodynamics. Entropy for 
a system 
of photons between the realistic metallic plates is calculated. It is shown
that in the approach of Refs.~\cite{28,29}, entropy is negative within the
separation range where the temperature correction to the Casimir energy 
density computed in \cite{28,29} is negative. It is proved also that both
in Refs.~\cite{28,29} on the one hand, and in \cite{35,36} on the other hand
the Nernst heat theorem, or the third law of thermodynamics, is violated.
Thus, the approaches of Refs.~\cite{28,29,35,36} are unacceptable from the 
thermodynamical point of view. As for the results of Refs.~\cite{30,31,33,34}
(for plasma model) and Refs.~\cite{32,33} (for Drude model), they are 
shown to be in agreement with the general principles of thermodynamics.

This paper is organized as follows. In Sec.~II the general expression 
for the entropy of a system of photons between metallic plates is
presented. 
Sec.~III contains the computational results for the entropy
in the framework of the Drude model using different approaches. In Sec.~IV
the analogical results  obtained in the framework of the plasma 
model are given. In Sec.~V the reader finds conclusions and discussion.

\section{Entropy for photons between plates made of real metal}

We consider two plane parallel plates made of real metal which are
in thermal equilibrium with a heat reservoir at some nonzero
temperature $T$. Let $a$ be the space separation between plates.
The modern derivation of the free energy per unit area for the
system under consideration is based on quantum field theory in the
Matsubara formulation and $\zeta$-regularization method.
The result is \cite{9}
\bes
&&
F_E(a,T)=\frac{k_B T}{4\pi}
\sum\limits_{l=-\infty}^{\infty}
\int_{0}^{\infty}
\kp d\kp
\label{1} \\
&&
\times
\left\{
\ln\left[1- r_{||}^2\left(\xi_l,\kp\right)
e^{-2aq_l}\right]+
\ln\left[1- r_{\bot}^2\left(\xi_l,\kp\right)
e^{-2aq_l}\right]
\right\},
\nn
\ees
\ni
where  
\beq
r_{||}^2(\xi_l,\kp)=\left[
\frac{\e (i\xi_l)q_l-k_l}{\e (i\xi_l)q_l+k_l}\right]^2\!\!,
\quad
r_{\bot}^2(\xi_l,\kp)=
\left(\frac{q_l-k_l}{q_l+k_l}\right)^2
\label{2}
\eeq
with
$q_l=(\xi_l^2/c^2+{\kp}^2)^{1/2}$, 
$k_l=[\e (i\xi_l)\xi_l^2/c^2+{\kp}^2]^{1/2}$
are the reflection coefficients for the modes 
corresponding to two different
polarizations. Here $\e$ is the frequency dependent dielectric 
permittivity of a plate material computed along imaginary
frequency axis at discrete Matsubara frequencies
$\xi_l=2\pi lk_BT/\hbar$ with 
$l=\ldots -2,\,-1,\,0,\,1,\,2\ldots\,$, $k_B$ is the Boltzmann
constant, and $\kp$ is the modulus of the wave vector component
in the plane of the plates.

The result (\ref{1}) is obtained by the solution of a
one-dimensional scattering problem on the axis perpendicular to
the plates. In fact, an electromagnetic wave which is coming
from the left in one semispace is scattered on the vacuum gap
between semispaces and there are reflected and transmitted
waves (see Refs.~\cite{9,32,33} for details).

It is important to keep in mind that the scattering problem
leading to Eqs.~(\ref{1}), (\ref{2}) has the definite solution only under
the requirement that
\beq
\lim\limits_{\xi\to 0}\xi^2{\e}(i\xi)=C\neq 0.
\label{3}
\eeq
\ni
If this is not the case (like for metals described by the Drude 
model or for dielectrics, see the next section) some additional 
conditions must be used to fix a solution of the scattering problem
at zero frequency. As an example, for dielectrics the results
(\ref{1}), (\ref{2}) are restated including the zero-frequency
contribution by using the unitarity condition and dispersion
relation. However, as to the case of the Drude model, describing
a medium with dissipation, the unitarity condition is not
applicable, and, therefore, a solution of the scattering problem at
zero frequency remains indefinite. Due to this fact, metals described
by the Drude model are outside of the application range of
formulas (\ref{1}), (\ref{2}) for the Casimir free energy density
at nonzero temperature.
The special prescription concerning the zero-frequency term of
Eq.~(\ref{1}) must be introduced in order that the dissipative
media could be described on the basis of this equation. This
prescription must be in accordance with the laws of thermodynamics
and other general physical requirements.

It is easily seen that Eqs.~(\ref{1}), (\ref{2}) lead to the famous 
Lifshitz formula \cite{19,27} for thermal Casimir force between
two semispaces which is obtained as
$F(a,T)=-\partial F_E(a,T)/\partial a$. Thus we arrived at the 
conclusion that the Drude metals are outside of the application
range of the Lifshitz formula at nonzero temperature and a
special prescription is needed in order to describe them in a
consistent way.

In the limit of zero temperature, the Casimir energy density of 
the zero-point electromagnetic oscillations is reobtained from 
Eq.~(\ref{1}) as
\bes
&&
E(a)=\frac{\hbar}{4\pi^2}
\int_{0}^{\infty}\!\!d\xi
\int_{0}^{\infty}\!\!
\kp d\kp
\left\{
\ln\left[1- r_{||}^2\left(\xi,\kp\right)
e^{-2aq_l}\right]\right.
\nn \\
&&\phantom{aaaaaaaaaaaaaaaaaa}
\left.+
\ln\left[1- r_{\bot}^2\left(\xi,\kp\right)
e^{-2aq_l}\right]
\right\}.
\label{4}
\ees
\ni
From (\ref{4}) the Lifshitz formula for the Casimir force at zero
temperature is obtained $F(a)=-\partial E(a)/\partial a$.
It is a nontrivial result that the Lifshitz formula at zero temperature
is applicable for nondissipative as well as for dissipative media.
This was demonstrated in Ref.~\cite{41} (see also \cite{21}) through
the consideration of a supplementary electrodynamical problem and is
explained by the fact that the only point $\xi=0$ 
which gives an important contribution to
the discrete sum (\ref{1}) does not contribute to the
integral (\ref{4}).

Note that the generally accepted terminology ``the Casimir energy
density and force at zero temperature'' is of some ambiguity.
It is really correct in the sense that the sum of the zero-point
energies (not the free energies) is computed. This terminology,
however, disregards the fact that there is a significant thermal
dependence of the energy density and force through the thermal
dependence of the dielectric permittivity (see the next section).
Because of this, when one calculates the Casimir energy density at, 
say, $T=300\,$K, the value of $\varepsilon(i\xi,T)$ is substituted
into Eq.~(\ref{4}), not $\varepsilon(i\xi,T=0)$ (see, e.g.,
Refs.~\cite{23,24,29,30}). Thus, instead of $E(a)$, a more exact
notation for the quantity (\ref{4}) would be $E_{T}(a)$.

If we multiply (\ref{1}) or (\ref{4}) by $2\pi R$, where $R\gg a$
is a radius of a sphere at a separation $a$ from the semispace, one
obtains the Casimir force in the configuration of a sphere near a plate
at a temperature $T$, or at zero temperature, respectively \cite{42}.

According to thermodynamics, the entropy
per unit area of the system under consideration is
\beq
S(a,T)=\frac{1}{T}\left[E_{T}(a)-F_E(a,T)\right],
\label{5}
\eeq
\ni
where $F_{E}(a,T)$ is given by Eq.~(\ref{1}) and $E_{T}(a)\equiv E(a)$
from Eq.~(\ref{4}).
In the next sections entropy is calculated for the plates made of real
metal as described by the Drude or plasma model dielectric functions.
In doing so, special attention is paid to the zero-frequency term
($l=0$) in Eq.~(\ref{1}) and to the fulfilment of Eq.~(\ref{3}).

\section{Entropy in the case of metallic plates described by the
Drude model}

It is common knowledge that the Drude dielectric function
\beq
\varepsilon_D(\omega)=1-\frac{\omega_p^2}{\omega(\omega+i\gamma)},
\label{6}
\eeq
\ni
where $\omega_p$ is the plasma frequency and $\gamma\ll\omega_p$
is the relaxation frequency, gives a good approximation of the dielectric 
properties for some metals, e.g., for aluminum. This approximation
was widely used in combination with the Lifshitz formula (\ref{4})
to calculate the finite conductivity corrections to the Casimir
force at zero temperature \cite{4,5,9,23,24}. To do so the Drude
dielectric function along the imaginary frequency axis was considered
\beq
\varepsilon_D(i\xi)=1+\frac{\omega_p^2}{\xi(\xi+\gamma)}.
\label{7}
\eeq

Before we proceed further, we note that the dielectric function (\ref{7})
violates the requirement (\ref{3}) for any nonzero value of the relaxation
frequency $\gamma$. Due to this, as discussed in the previous section,
the Drude metals are outside of the application range of the Lifshitz
formula (\ref{1}) at nonzero temperature with an unmodified zero-frequency
term. If, nevertheless, one substitutes Eq.~(\ref{7}) into Eq.~(\ref{1}),
as was done, e.g., in Refs.~\cite{6,28,29}, several questionable results
follow which are in contradiction with the limiting cases of metal
described by the plasma model (see the next section) and of an ideal metal
(see Introduction and a detailed discussion in Refs.~\cite{32,33,34}).
Using the dielectric function of Eq.~(\ref{7}), the values of
reflection coefficients (\ref{2}) at zero frequency are the following
\beq
r_{||}^2(0,k_{\bot})=1,
\quad
r_{\bot}^2(0,k_{\bot})=0.
\label{8}
\eeq
\ni
From the formal point of view, some troubles are connected with the second
of Eqs.~(\ref{8}) because for an ideal metal $r_{\bot}^2(0,k_{\bot})=1$.

In order to present the crucial argument against the substitution
of the Drude dielectric function (\ref{7}) into the unmodified
Lifshitz formula (\ref{1}) at nonzero temperature, we calculate the
entropy per unit area given by Eq.~(\ref{5}) in the framework of
the Drude model. Thus, as an example, consider $Al$ plates with
the parameters \cite{43}
\bes
&&
\omega_p\approx 12.5\,\mbox{eV}\approx 1.9\times 10^{16}\,\mbox{rad/s},
\label{8a} \\
&&
\gamma\approx 0.063\,\mbox{eV}\approx 9.6\times 10^{13}\,\mbox{rad/s}.
\nn
\ees
\ni
In Fig.~1 the computational results for the entropy are presented at
$T=300\,$K as a function of the separation between the plates. 
Note that the plasma frequency practically does not depend on
temperature. As to the value of the relaxation frequency from
Eq.~(\ref{8a}), it is given for the temperature under consideration.
It is seen from the figure 
that the entropy is negative within a wide separation range
that it not acceptable from a thermodynamical point of view.
The separation interval $0<a<4.1\,\mu$m, where the entropy is
negative, coincides with the interval where the negative temperature
corrections arise in the approach used in \cite{28,29} (see the detailed
discussion in \cite{32}). Negative temperature corrections are in
conflict with the evident physical arguments (with increase of
temperature the number of photons in the modes and thereby force modulus 
should increase). Here we show that they are also in contradiction with
the general physical principles.

As noted in the Introduction, except of the immediate application of the
Lifshitz formula in combination with the Drude model \cite{28,29},
different prescriptions were proposed in literature modifying
the zero-frequency term of this formula in the case of real metals.
In \cite{35,36} it was postulated that
\beq
r_{||}^2(0,k_{\bot})=
r_{\bot}^2(0,k_{\bot})=1,
\label{9}
\eeq
\ni
as in the case of an ideal metal. This prescription was criticized on physical
grounds in \cite{32}.

In Ref.~\cite{32} the other prescription was proposed which is the
generalization of the prescription of Ref.~\cite{20}, formulated for
an ideal metal. According to \cite{32} in the framework of the Drude model
the reflection coefficients at zero frequency are
\beq
r_{||}^2(0,k_{\bot})=1,
\quad
r_{\bot}^2(0,k_{\bot})=
\left(\frac{ck_{\bot}-\sqrt{\frac{\omega_p^2ck_{\bot}}{ck_{\bot}+
\gamma}+c^2k_{\bot}^2}}{ck_{\bot}+\sqrt{\frac{\omega_p^2ck_{\bot}}{ck_{\bot}+
\gamma}+c^2k_{\bot}^2}}\right)^2.
\label{10}
\eeq
\ni
As shown in \cite{32}, prescription (\ref{10}) leads to wholly
satisfactory results.

In order to test all the above approaches for conformity to the general
principles of thermodynamics, we find the dependence of the entropy 
(\ref{5}) on temperature at some fixed plate separation, say,
$a=2\,\mu$m. To accomplish this, one should take into consideration
that except of the evident dependence of Eqs.~(\ref{1}) and (\ref{5})
on temperature there is the aforementioned significant thermal dependence of
$\varepsilon_D$ given by Eq.~(\ref{7}) through the relaxation parameter
$\gamma=\gamma(T)$ (coinciding with the thermal dependence of
resistance). The dependence $\gamma(T)$ is linear at temperatures
higher than $0.25T^{\ast}$, where $T^{\ast}$ is the Debye temperature
($T^{\ast}=428\,$K for $Al$ \cite{44}). At lowest temperatures $\gamma(T)$
follows the power law $T^n$ ($n=2-5$ depending on the metal). The complete
dependence of the nondimensional normalized quantity
$\tilde{\gamma}(T)\equiv 2a\gamma(T)/c$ on the temperature is plotted
in Fig.~2 by the use of tabulated data for $Al$ \cite{44} (here we
neglect the small residual resistivity caused by the scattering of  
electron waves by static defects that is outside of the frameworks of the
Drude model \cite{45}). 

Now we are in a position to calculate the dependence of entropy on 
temperature for all the above approaches. Calculation was performed  
using Eq.~(\ref{5}) and also Eqs.~(\ref{1}), (\ref{4}), (\ref{7})
and Fig.~2. The results are presented in Fig.~3. The long-dashed curve
is obtained on the basis of the Lifshitz formula with an unmodified 
zero-frequency term (i.e., Eq.~(\ref{8}) was used for the reflection
coefficients at zero frequency). Remind that this approach was
exploited in Refs.~\cite{28,29}. The short-dashed curve is calculated
with the modification of the zero-frequency term of the Lifshitz formula
in accordance with Eq.~(\ref{9}) (approach of Refs.~\cite{35,36}).
The solid curve is calculated with the modification of the zero-frequency
term according to Eq.~(\ref{10}) suggested in Ref.~\cite{32}. Note that
all the above approaches differ by the value of the zero-frequency term
only for perpendicular polarization.

As is quite clear from Fig.~3, for both long-dashed and short-dashed
curves the values of entropy at zero temperature are not equal to zero.
In the approach used in \cite{28,29} 
$S_1(0)=-0.5\,\mbox{MeV}\,\mbox{m}^{-2}\,\mbox{K}^{-1}$,
and in the approach used in \cite{35,36} 
$S_2(0)=0.016\,\mbox{MeV}\,\mbox{m}^{-2}\,\mbox{K}^{-1}$.
In both cases the value of $S(0)$ depends on the parameters of the
system under consideration (like the separation between the plates
and the plasma frequency) that is in manifest contradiction with the
third law of thermodynamics (the Nernst heat theorem \cite{46,47}).
It is notable that the entropy density given by the long-dashed curve
is negative in a wide temperature range (compare with Fig.~1 where the 
result at a fixed temperature is presented). It is easily shown that the
values of entropy density at zero temperature, given by the dashed curves,
are related by 
\beq
S_2(0)-S_1(0)=\frac{k_B\zeta(3)}{16\pi a^2},
\label{11}
\eeq
\ni
where $\zeta(3)$ is the Riemann zeta function. The right-hand side of
Eq.~(\ref{11}) is equal to one-half of the coefficient near temperature 
in the zero-frequency term of the Lifshitz formula for an ideal metal.
This is because in \cite{35,36} the same values (\ref{9}) for the
reflection coefficients at zero frequency were postulated as for an ideal
metal, whereas in \cite{28,29}, according to (\ref{8}), one-half of
the result for an ideal metal was used. In fact the correct result for 
a real metal lies in between of these two possibilities.

Contrary to the dashed curves, for the solid curve (approach of 
Ref.~\cite{32}), $S(0)=0$ and therefore, in 
accordance with the laws of thermodynamics.

\section{Entropy in the case of metallic plates described by the
plasma model}

For the free-electron plasma model the dependence of dielectric
function on the frequency is given by
\beq
\varepsilon_p(\omega)=1-\frac{\omega_p^2}{\omega^2},
\quad
\varepsilon_p(i\xi)=1+\frac{\omega_p^2}{\xi^2}.
\label{12}
\eeq
\ni
This dependence was widely used to calculate the finite conductivity
corrections to the Casimir force at the separations of order $1\,\mu$m
\cite{19,20,21,22}. In Refs.~\cite{30,31} it was applied to compute
the effect of nonzero temperature and finite conductivity in the framework
of the Lifshitz formula (\ref{1}).
Note that the plasma dielectric function practically does not depend
on temperature.

The preference for the plasma model as compared with the Drude one
is the fulfilment of condition (\ref{3}) with the dielectric function
(\ref{12}). As a consequence, the scattering problem underlying the
Lifshitz theory has a definite solution leading to Eqs.~(\ref{1}),
(\ref{2}), (\ref{4}) including the zero-frequency contribution.
The reflection coefficients (\ref{2}) at zero frequency in the framework
of the plasma model take the form
\beq
r_{||}^2(0,k_{\bot})=1,
\quad
r_{\bot}^2(0,k_{\bot})=
\left(\frac{ck_{\bot}-\sqrt{\omega_p^2+
c^2k_{\bot}^2}}{ck_{\bot}+\sqrt{\omega_p^2+
c^2k_{\bot}^2}}\right)^2.
\label{13}
\eeq
\ni
Note that Eq.~(\ref{13}) can be obtained from Eq.~(\ref{10}) when the
relaxation frequency goes to zero. Because of this, in the approach of
\cite{32} the results for the plasma model are obtainable from the
results for the Drude model in a limitimg case $\gamma\to 0$
(the results obtained in \cite{28,29} by the use of the Drude model
on the basis of Eq.~(\ref{8}) have no smooth connection with a plasma
model approach; the plasma model by itself is not considered in
\cite{28,29}). In the alternative approach \cite{35,36}, as distinct
from both \cite{30,31,32,33} and \cite{28,29}, the conditions (\ref{9})
are postulated in the framework of both Drude and plasma models.

Now let us find the dependence of the entropy density on temperature
in the framework of the plasma model on the basis of different
approaches. As an example, $Al$ plates are used once more at a separation
$a=2\,\mu$m. The value of the plasma frequency for $Al$ is given by
Eq.~(\ref{8a}) \cite{43}. Calculations were performed using
Eqs.~(\ref{1}), (\ref{4}), (\ref{5}), (\ref{7}) and (\ref{12}). 
The results are presented in Fig.~4. The dashed curve is obtained in
the framework of Refs.~\cite{35,36} using prescription (\ref{9}).
The solid curve is calculated on the basis of Refs.~\cite{30,31,32,33}
with no modification of the Lifshitz formula, i.e. with the zero-frequency
reflection coefficients (\ref{13}).

As is obvious from Fig.~4, for the dashed curve the value of entropy
at zero temperature
$\tilde{S}(0)=S_2(0)=0.016\,\mbox{MeV}\,\mbox{m}^{-2}\,\mbox{K}^{-1}$
and is not equal to zero. It depends on the parameters of the system 
that is in contradiction with the third law of thermodynamics.
In contrast to this, for the solid curve obtained on the basis of
the fundamental Lifshitz formula, $S(0)=0$ as it must be from the third
law of thermodynamics.

In the framework of the plasma model it is not difficult to obtain the
analytical expression for the entropy at low temperatures
$k_BT\ll k_BT_{eff}\equiv\hbar c/(2a)$. For this purpose the perturbation
expansion of the Casimir energy and free energy in powes of two small
parameters $\delta_0/a=c/(a\omega_p)$ and $T/T_{eff}$ can be used
(see Refs.~\cite{22,31,33} where these expansions are presented in details;
they are applicable for separations $a\geq\lambda_p=2\pi c/\omega_P$
where $\lambda_p$ is the effective plasma wavelength). Substituting the
mentioned perturbation expansions into Eq.~(\ref{5}), one obtains
\bes
&&
S(a,T)=\frac{k_B\zeta(3)}{8\pi a^2}
\left(\frac{T}{T_{eff}}\right)^2
\left\{
\vphantom{\left[\frac{2\pi^2}{45\zeta(3)}\frac{T}{T_{eff}}\right]}
1-\frac{\pi^3}{45\zeta(3)}\frac{T}{T_{eff}}\right.
\label{14} \\
&&
\phantom{aaa}
+\left.
2\frac{\delta_0}{a}\left[1-
\frac{2\pi^3}{45\zeta(3)}\frac{T}{T_{eff}}\right]\right\}.
\nn
\ees
\ni
Here the terms up to $(T/T_{eff})^3$ were included. The powers of
$\delta_0/a$ higher than one are contained only with powers
$(T/T_{eff})^n$, $n>3$.
This analytical expression corresponds to the solid curve in Fig.4.
It is seen from Eq.~(\ref{14}) that the entropy approaches zero as
the second power of temperature in the framework of the unmodified
Lifshitz formula.

On the basis of the approach proposed in Refs.~\cite{35,36} the perturbation
expansion of entropy is given by
\beq
\tilde{S}(a,T)=\frac{k_B\zeta(3)}{4\pi a^2}\frac{\delta_0}{a}
\left(1-3\frac{\delta_0}{a}\right)+S(a,T),
\label{15}
\eeq
\ni
where
$S(a,T)$ is expressed by Eq.~(\ref{14}). The first contribution
in the right-hand side of Eq.~(\ref{15}) is the value of entropy
at zero temperature
\beq
\tilde{S}(a,0)=\frac{k_B\zeta(3)}{4\pi a^2}\frac{\delta_0}{a}
\left(1-3\frac{\delta_0}{a}\right)\neq 0.
\label{16}
\eeq
\ni
At $a=2\,\mu$m one obtains from Eq.~(\ref{16}) the above value of 
$S_2(0)$.
$\tilde{S}(a,0)$ depends on both the separation between the plates
$a$ and the penetration depth of the electromagnetic oscillations into the
plate material $\delta_0$ in contradiction with the third law of
thermodynamics \cite{46,47}. 

\section{Conclusion and discussion}

In the foregoing we have considered the thermodynamical aspects of
the Casimir force acting between real metals at nonzero temperature.
The necessity of considering these aspects stems from the
controversial results obtained by different authors (see
Refs.~\cite{28,29,30,31,32,33,34,35,36}) and continuing polemic
\cite{37,38,39,40}. The further importance to this problem is
added by the rapid progress in experiment on measuring the Casimir
force. At the moment, there is a contradiction between the
experimental results of \cite{1} and the theoretical approach
of \cite{28,29} which leads to large negative temperature
corrections at separation of about $1\,\mu$m (see the discussion
in \cite{37,40}). On the other hand, the experimental results of
\cite{2,3,4,5} do not agree with the computations of \cite{35,36}
that lead to large (although positive) linear in temperature
corrections to the Casimir force at separations of about 
100\,nm (see \cite{32}).

Bearing in mind that there are many influential factors in so
precise experiments, it is highly desirable to offer some decisive
theoretical arguments providing a way to give preference to one
of the theoretical approaches. As shown above, thermodynamics
gives the possibility to make a selection and to reject the
approaches that are not in accordance with the most fundamental
physical principles.

In the present paper we calculated the entropy density for photons
between two parallel plates made of real metals described by the
Drude or plasma models. It is shown that in the approach of
Refs.~\cite{28,29}, based on a direct application of the 
unmodified Lifshitz formula in the case of Drude metals, entropy
takes negative values in a wide range of related parameters.
The value of entropy density at zero temperature, as given by
the approach of \cite{28,29}, is shown to be nonzero and dependent
on the parameters of the system under consideration in contradiction
with the Nernst heat theorem. These unacceptable properties
of entropy confirm the conclusion of Refs.~\cite{32,33} that the
dissipative metals described by the complex dielectric permittivity
of real frequency are outside of the application range of the
Lifshitz formula at nonzero temperature. To describe the thermal
Casimir force for such metals a special prescription should be
adopted modifying the zero-frequency term of the Lifshitz formula.

One prescription of this kind (the same as was proposed in \cite{20} 
for the case of ideal metal) was suggested in Refs.~\cite{35,36}.
We show that although the entropy density in the approach of 
\cite{35,36} is positive, the value of entropy at zero temperature 
is not equal to zero and depends on the parameters of the system.
Thus, this approach is also in contradiction with the third law
of thermodynamics.

One more prescription for the zero-frequency term of the Lifshitz
formula was proposed in \cite{32}. It is the generalization of the
receipt of \cite{20} for the case of real metals.
We show that for the prescription suggested in \cite{32}, entropy
is non-negative at all temperatures and takes zero value
at zero temperature. Hence the prescription of Ref.~\cite{32}
is in agreement with the general principles of thermodynamics.

In this paper we report also the results of the computation
of the entropy in the framework of the plasma model. The plasma
model does not take dissipation into account.
It belongs to the application range of the Lifshitz formula.
It is shown that the application of the unmodified Lifshitz
formula in combination with the plasma model leads to wholly
satisfactory results: the entropy density is positive and takes
zero value at zero temperature.
The application of the modified Lifshitz formula, as in 
Refs.~\cite{35,36}, leads to the violation of the third law
of thermodynamics.

To conclude, the approach of Refs.~\cite{28,29} on the one hand 
and of Refs.~\cite{35,36} on the other hand must be rejected as 
they are in contradiction with the general principles of 
thermodynamics. In the case of the Drude metals only the approach
of Ref.~\cite{32} fits thermodynamical requirements. It is also
in accordance with the present experimental results. As to the
case of the plasma metals with no account of dissipation, the
unmodified Lifshitz formula is applicable and the results of
Refs.~\cite{30,31,32,33,34} are in agreement between themselves 
and with the general principles of thermodynamics.

\section*{ACKNOWLEDGMENTS}

The authors are grateful to CNPq for partial financial
support.


\newpage
\widetext
\begin{figure}[h]
\vspace*{-7cm}
\epsfxsize=20cm\centerline{\epsffile{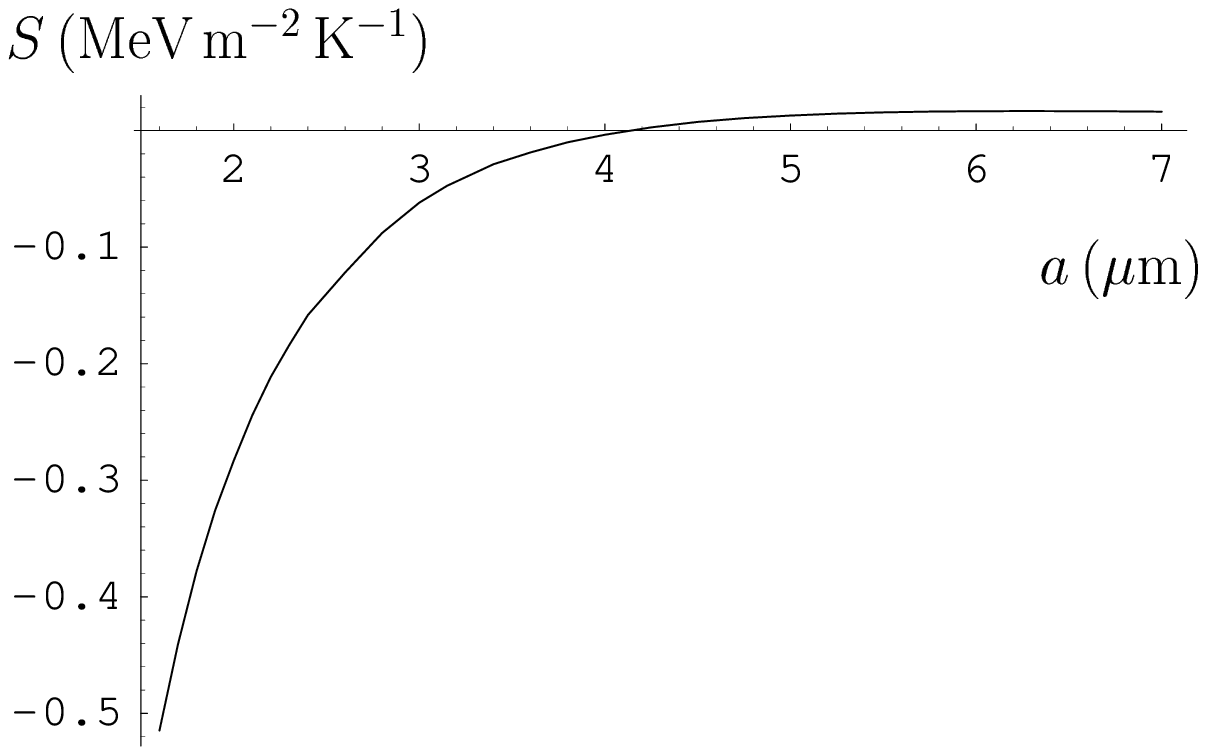}}
\vspace*{-8cm}
\caption{Entropy of photons between $Al$ plates described by 
the Drude model at $T=300\,$K as a function of space separation
computed using the approach of Refs.~[30,31]. 
}
\end{figure}
\newpage
\begin{figure}[h]
\vspace*{-7cm}
\epsfxsize=20cm\centerline{\epsffile{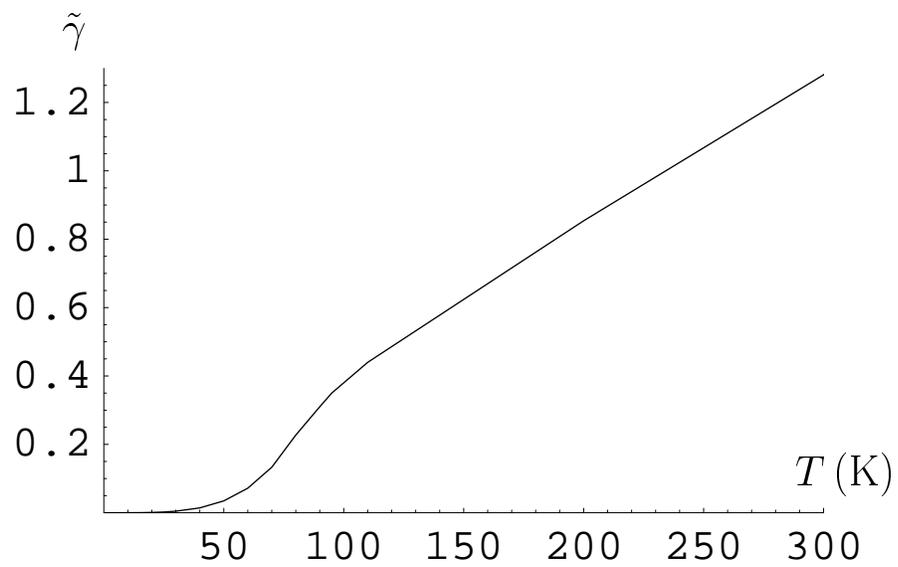}}
\vspace*{-8cm}
\caption{Dimensionless relaxation frequency of $Al$ as 
a function of temperature.
}
\end{figure}
\newpage
\begin{figure}[h]
\vspace*{-7cm}
\epsfxsize=20cm\centerline{\epsffile{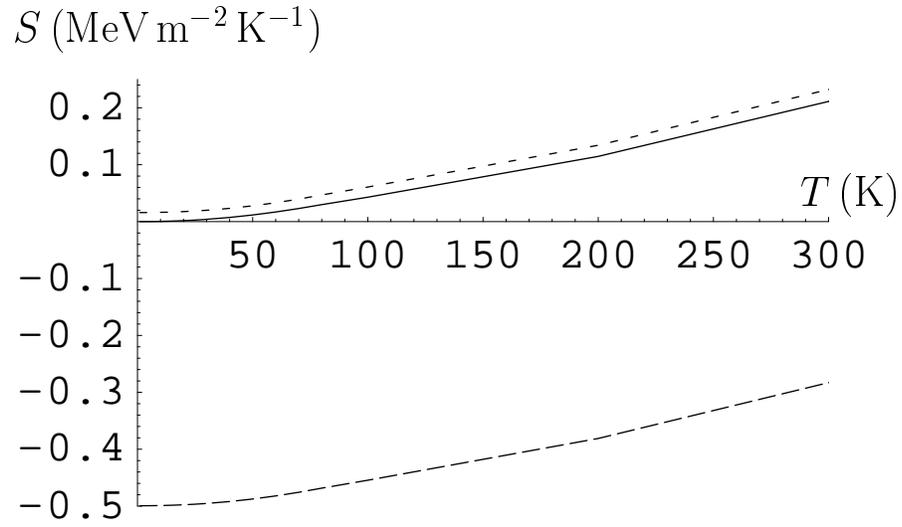}}
\vspace*{-8cm}
\caption{Entropy of photons between $Al$ plates described by 
the Drude model at a separation  $a=2\,\mu$m as a function of
the temperature. Long-dashed curve was 
computed using the approach of Refs.~[30,31], the short-dashed curve 
was obtained with the
approach of [37,38], and the solid curve using the approach of [34]. 
}
\end{figure}
\newpage
\begin{figure}[h]
\vspace*{-7cm}
\epsfxsize=20cm\centerline{\epsffile{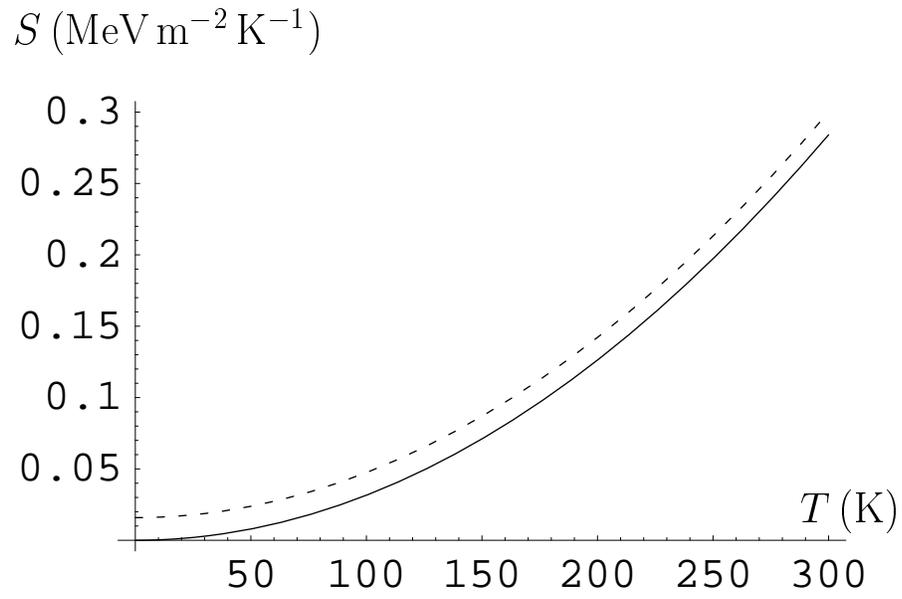}}
\vspace*{-8cm}
\caption{Entropy of photons between $Al$ plates described by 
the plasma model at a separation  $a=2\,\mu$m as a function of
the temperature. Dashed curve was 
computed using the approach of [37,38] 
and the solid curve was obtained with the approach of [32--36]. 
}
\end{figure}
\end{document}